# Violation of Bell Inequalities: Mapping the Conceptual Implications

**Brian Drummond**

Edinburgh, Scotland. *E-mail:* *drummond.work@phonecoop.coop*

**Abstract:** This short article concentrates on the conceptual aspects of the violation of Bell inequalities, and acts as a map to the 265 cited references. The article outlines (a) relevant characteristics of quantum mechanics, such as statistical balance and entanglement, (b) the thinking that led to the derivation of the original Bell inequality, and (c) the range of claimed implications, including realism, locality and others which attract less attention. The main conclusion is that violation of Bell inequalities appears to have some implications for the nature of physical reality, but that none of these are definite. The violations constrain possible prequantum (underlying) theories, but do not rule out the possibility that such theories might reconcile at least one understanding of locality and realism to quantum mechanical predictions. Violation might reflect, at least partly, failure to acknowledge the contextuality of quantum mechanics, or that data from different probability spaces have been inappropriately combined. Many claims that there are definite implications reflect one or more of (i) imprecise non-mathematical language, (ii) assumptions inappropriate in quantum mechanics, (iii) inadequate treatment of measurement statistics and (iv) underlying philosophical assumptions.

## 1. Introduction and Overview (Area Mapped and Mapping Methods)

Concepts are an important part of physics [1, § 2][2][3, § 1.2][4, § 1][5, p. 82][6,7]. This article, therefore, concentrates on the conceptual aspects of the violation of Bell inequalities. To do so, it builds on the analysis in *Understanding quantum mechanics: a review and synthesis in precise language* [3] (the *2019 review*) and, like that review, uses only non-mathematical language. The use of such language "can enhance understanding of the mathematics of quantum mechanics, for at least some users" [3, § 1.3]. The use of non-mathematical language raises the question: which non-mathematical language? This article is (a) written in English, and (b) based on literature written in, or translated into, English. This article therefore reflects only limited "epistemic diversity" [8].

Subject to that language constraint, this article aims to refer to enough of the relevant literature to be representative of the current state of the subject. This article is, to the relevant literature, what a map is to the territory it represents. [3, § 1.4]

Underlying philosophical prejudices, sometimes unconscious, will very often affect how the violation of Bell inequalities is discussed. Given the potential need to modify pre-quantum mechanical concepts in this context, it is important to consider what these prejudices might be, and to make assumptions explicit [3, § 1.1][9, p. 117][10, p. 204]. Some of the content of this article relates to wider discussions on the themes of reality, spacetime, probability and determinism. Underlying assumptions in these areas, and the intended meanings of related ambiguous terms, are as in the 2019 review. The first part of that review outlined these assumptions with reference to the relevant literature [3, § 1.5–§ 1.9].

Some words used in discussing the inequalities can take a variety of meanings. Imprecise non-mathematical language can make assessing the implications of the violation of Bell's inequalities harder than it needs to be [3, § 1.1]. An equivalent concern has long been acknowledged in wider physics [5, pp. 74–75], and has more recently been highlighted in wider quantum mechanics [6, § 26.2]. The final part of the 2019 review is a glossary of intended meanings for many elements of the non-mathematical language used in that review [3, § 8]. The intended meanings for such language are as in the 2019 review, subject to the three additions and one amendment set out in Part 5 below. These meanings may differ from the meanings intended by other authors, when they use the same words [3, § 1.4].

Part 2 outlines (a) relevant characteristics of quantum mechanics, such as statistical balance and entanglement, and (b) the thinking that led to the derivation of the original Bell inequality. Part 3 surveys the range of claimed implications, including those which attract less attention, and specifically considers realism and locality. The main conclusion, in Part 4, is that violation of Bell inequalities appears to have some implications for the nature of physical reality, but that none of these are definite. Many claims that there are definite implications reflect one or more of (a) imprecise non-mathematical language, (b) assumptions inappropriate in quantum mechanics, (c) inadequate treatment of measurement statistics and (d) underlying philosophical assumptions.

## 2. The Bell Inequalities in Context (Small Scale Map of a Large Area)

### 2.1. Quantum Mechanics: Statistics and Probabilities

> Quantum mechanics prescribes (specifies in advance) some aspects of expected future events relating to physical systems, in a range of possible situations [11][12, § 4.5]. Prescriptions (advance specifications) are made, collectively, as probability distributions [3, § 2.1].

As a result, "verifying the prescriptions of quantum mechanics is almost always statistical [13, § 4.2.3, § 6.4][14, ch. 9][15–18][19, pp. 206, 210][20, p. 99]" [3, § 2.1][21, p. 224]. The concept of statistical ensemble underlies the concept of quantum mechanical state, a core element of the quantum mechanical formalism [3, § 3.2][21, p. 224]. Also central to quantum mechanics is the fact that the prescriptions inherent in any quantum mechanical state will cover a range of different, often mutually exclusive, physical contexts [3, § 3.2][22, p. 224][23][24, p. 2]. Even quantum mechanical analyses that, in theory, seem to depend neither on inequalities nor statistics, remain inherently statistical in that they are quantum mechanical [25].

Appropriate analysis of probability, therefore, is central to quantum mechanics in general [3, § 2.1][26–30], and is crucial, in particular, to assessing the implications of violation of Bell inequalities [31,32][33, § 5][34, p. 5][35, p. 1603]. For example, Bell inequalities reflect, or even reproduce, the Boole inequality in probability theory. Over 100 years before Bell's work, Boole had identified that violation of this (Boole) inequality implies that, for a system of three random variables, a joint probability distribution does not exist [26, § 4.1][36, § 2][37, § 4][38, p. 167][39, § 2][40, § 1, § 4].

Likewise, the non-mathematical language used to deal with probability requires care [3, § 1.7, § 1.8][41,42]. Again, this is particularly true in assessing the implications of violation of Bell inequalities [43].

*2.2. Statistical Balance in Quantum Mechanics*

The 2019 review highlights the concept of statistical balance [3, § 2.2]: "for some combinations of measurement types, the observed statistics indicate that the collective response of (what are taken to be) identically prepared systems to differing measurement types is not at all straightforward". *Statistical balance* refers to this balanced collective response, to differing measurement types [44, Part 3], which features even in the analysis of systems which are single (no subsystems) and simple (no structure) [3, § 2.2][45, § 2][46, § 8][47, § 4][48, pp. 10–11].

This article will also use *statistical balance* as a convenient term for this core feature of quantum mechanics: the fact that, "for any given measurement type, in a series of measurement events, the outcomes (collectively) give statistics consistent with the prescribed probability" [3, § 2.2]. The choice of this term derives from the fact that there is a sense of balance in the statistics. For example, there is a balance between the outcomes of individual runs when the overall result of the measurement is in line with a prescribed probability which is neither 0 nor 1. This is true even when it is demonstrable that no definite value for the measured property can be attributed to individual members of the originally prepared ensemble [3, § 2.2][49, § 1.1.2]. The balance is among measurement event outcomes which happen sequentially in time. This article will refer to this instance of statistical balance as *two-measurement-type statistical-balance-in-time*.

The word balance has some connotations which might make it seem inappropriate in this context. For example *balance* might suggest a feature with a known underlying explanation, or even a feature in which two things are equally balanced. Neither connotation is intended by the choice of this term. Indeed statistical balance characterizes a series of measurement event outcomes, with none of them being explicable by reference to pre-measurement underlying properties. The fact that the overall result of the measurement is predictable is very hard to explain without significant modification of pre-quantum mechanical concepts.

The 2019 review mentioned four suggested approaches to explaining statistical balance within an ensemble. The last of these involved recognising the concept of potentia as a feature of independent reality, and continues to be developed [50–54].

*2.3. Entanglement: Statistically Balanced Subsystem Outcomes*

When systems are considered together in quantum mechanics, any separate states for the systems are replaced by a new one for the composite system [46, § 15]. The term entangled state can be used to refer to a mixed composite system state [55, p. 2][56, p. 5], but this article will follow the 2019 review in limiting the use of the term *entangled state* to pure states, such as superpositions of composite system pure states [3, § 3.6, § 8][57, p. 149]. Such pure entangled states are, however, recognised to be an idealisation [3, § 3.6][58].

"Like all quantum mechanical states, entangled states feature a statistical balance in collective outcomes, among ensembles for differing measurements on the ensemble to which they relate" [3, § 3.6]. What distinguishes entangled states from non-entangled states is statistical balance among collective outcomes, for differing measurement types, on far-apart subsystems [3, § 3.6][59]. This article will refer to the latter instance of statistical balance as *two-measurement-type statistical-balance-in-space-and-time*. Entanglement can arise irrespective of the relationship between the two measurement types [60].

### 2.4. Background to the Original Bell Inequality

The Einstein-Podolsky-Rosen thought experiment [61] has been subject to extensive analysis [62][3, § 5.2]. When an observer, O2, carries out a measurement on system 2, of a pair of systems, it appears that O2 can change the state representing system 1. Schrödinger informally described this as O2 steering, or piloting, system 1 [63]. More recent and precise analysis highlights that the pair of systems must be treated as a composite system represented by an entangled state [3, § 5.2][64, § 3.2][65,66][67, § 3.4][68, § 5].

Despite frequent reference to correlations and causation in this context [69], such an approach is inappropriate [3, § 5.3][70, p. 3]. The Einstein-Podolsky-Rosen pair is reflecting features which pervade quantum mechanics [45,47], features which are also seen in non-composite systems [3, § 5.3][13, § 6.4][46, § 10, § 11][71, § V.F][72].

Building on the Einstein-Podolsky-Rosen analysis, Bell explored some possible implications of the quantum mechanical analysis of a composite system entangled state [73], in which "the result of measuring any chosen spin component for one subsystem, can be predicted by first measuring the same component for the other" [3, § 5.4]. Bell first hypothesised that, when the subsystems are far apart, the choice of which component of spin is to be measured on one subsystem does not influence the result of a measurement on the other. Bell then took this to imply that the result of the second measurement was predetermined, which suggests that it might be possible to specify the system more completely than is done by the quantum mechanical state. For one such possible specification, Bell derived a resulting inequality, and showed that it is violated by the predictions of quantum mechanics [73].

Since Bell's original paper, Bell's name has been associated with many other inequalities relating to composite systems [74, Appendix], and more continue to be developed [31]. Most of this article is potentially applicable to any of these many inequalities. On that basis, the phrase *Bell inequalities* in this article will mean any of the inequalities classified in the 2014 review *Bell nonlocality* [74, Appendix].

## 3. Conceptual Implications of Violation of Bell Inequalities (Larger Scale Map of a Smaller Area)

### 3.1. Bell Inequalities: Statistical Balance

Bell inequalities involve quantum mechanical analysis of widely-extended composite systems. In assessing the implications of violation of Bell inequalities, it is important to remember known features quantum mechanical analysis of systems which are non-composite (no subsystems) and simple (no extended structure). In particular, as noted in Sect. 2.2 above, the prescribed statistics of measurement event outcomes for non-composite simple systems reflect two-measurement-type statistical-balance-in-time: the overall result of a measurement is often in line with a prescribed probability which is neither 0 nor 1, despite the fact that no definite value of the relevant property can be attributed to individual members of the originally prepared ensemble. If this is true for a collection of measurement event outcomes which happen sequentially in time for two-measurement-types, it should not be surprising to discover a similar statistical balance among measurement event outcomes which happen sequentially in space and time for two measurement-types (two-measurement-type statistical-balance-in-space-and-time).

At least some of the conceptual confusion which often features in discussion of Bell inequalities results from a failure to apply the caution which should be motivated by the statistical balance which pervades quantum mechanics. In particular "[t]here seems to be a widespread, implicit acceptance that explaining this balance is not part of quantum mechanics" [3, § 2.2][45,47]. In marked contrast, there seems to be a widespread, explicit acceptance that explaining violation of Bell inequalities *is* part of quantum mechanics.

The inconsistency of this approach to explanation has recently been highlighted in a startling way. An approach, similar to that taken to quantum mechanical analysis of composite systems and Bell inequalities, can be taken to analysing two differing measurement types being applied consecutively to a non-composite system. If this is done, then it is possible to infer characteristics of a type typically associated with quantum mechanical analysis of composite systems [75] (although the conventional analysis of the particular set-up considered in that analysis is "notoriously counter-intuitive" and potentially in need of revision [76]). In a similar way, the conclusions drawn by Einstein-Podolsky-Rosen by considering two measurement types for a composite system could equally well have been drawn by considering a single measurement type for a composite system [77, § 3.3].

Thus, violation of Bell inequalities reflects, at least in part, the two-measurement-type statistical-balance-in-time that would be evident in a single non-composite system [78, p. 643][79]. It has been suggested that this is the sole or main implication and, therefore, that discussion of nonlocality is misguided [80–82], but some aspects of this view are currently being challenged [83].

In the absence of a theory to explain the pervasive statistical balance, it would be surprising if the particular balance reflected in violation of Bell inequalities could be explained. On that basis, the violation of Bell inequalities seems a particularly inappropriate place to start to draw conclusions on questions such as realism or locality. Conversely, the plausibility of a possible explanation for the pervasive statistical balance, such as a conservation principle operating on average for an ensemble, can be tested by reference to the Bell inequalities [84].

### 3.2. Bell Inequalities: Locality and Realism

Some analyse Bell inequalities in terms of locality [85, § 8.6][86–88], causality [85, § 8.7][89,90] and local causality [91–95]. It is challenging to achieve appropriate definitions of these terms [56, pp. 7–8][70, p. 3][96–98]. For example there may be value in distinguishing between: the apparent nonlocality suggested by violation of Bell inequalities; and the *steering* effect arising in the Einstein-Podolsky-Rosen context [56,99–107]. Conceptual analysis of these terms and definitions is also ongoing. For example, there can be a benefit in clarifying the concept of separability [56, p. 7][108], distinguishing separability from locality [67, § 3.3][87, ch. 8], or identifying to what extent non-separability arises in classical analysis [109, § 3][110, p. 14].

Some suggest that violation of Bell inequalities rules out local realism [111, pp. 487–488], and so forces a choice between two alternatives: realism or locality, but views differ of the meaning of these terms [37,40,87,91,112–118], and there is no clear consensus in this area [119–122]. Careful conceptual analysis of the term realism is needed in the context of quantum mechanics, as is clarity on which of the many meanings of the word is meant in any discussion of Bell inequalities [24,

pp. 5–6][123][124, § 3][125, § 5.2].

A striking example of lack of consensus on the implications of the inequalities is the ongoing debate on (a) what Bell took the Einstein-Podolsky-Rosen analysis to have demonstrated, and (b) whether Bell intended his original analysis to complement the Einstein-Podolsky-Rosen analysis, or to stand on its own. If the Einstein-Podolsky-Rosen analysis is seen to demonstrate that locality necessarily implies predetermined results, and if Bell's original analysis is seen to demonstrate that predetermined results are necessarily inconsistent with quantum mechanics, then locality is necessarily inconsistent with quantum mechanics [88, § 4.5.2][125, § 5.2][126, § 3.9][127, § 13.2]. Although this argument is logically correct, both of the premises are open to challenge. This debate is prolonged by three other factors: (i) the less than clear presentation of the original Einstein-Podolsky-Rosen analysis [127, § 13.2][128, § 6.3]; (ii) the fact that Bell's original analysis makes only very brief reference to the Einstein-Podolsky-Rosen analysis [125, § 5.7]; and (iii) disagreement over the extent to which Bell's subsequent writings reflected new results and thinking, or revised expressions of the original result and thinking [128–130].

Several projects are exploring prequantum theories, which might underlie quantum mechanics. Some of these projects aim to reconcile at least one understanding of locality and realism to quantum mechanical predictions, including Bell inequality violations [24, pp. 3–4][40][131, § 3.6][132–135][136, § 1]. Other projects are exploring possible prequantum theories which allow for the possibility of some form of nonlocality [64, pp. 51, 54][137]. Bell analyses constrain both groups of prequantum theories [138], but do not necessarily rule them out [3, § 7][139, § 2][140–142][143, pp. 136–137]. Recent attempts to rule out such theories are also unsuccessful [144], perhaps necessarily so [145, § 5][146,147][148, § 8].

Many views of realism arising in discussion of Bell inequalities seem inconsistent with what are otherwise accepted as core features of quantum theory [3, § 5.6][13, § 9.1.3][16, § 1][72,149–151][152, § 2.2][153, p. 7][154, pp. 153–154, 158–159]. Similar inconsistency (between the combinations of assumptions used in the theorems and relevant features of quantum theory) continues to be a feature of at least some theorems seeking to draw conclusions about the nature of mind-independent reality [155–157].

Among those who challenge the assumptions underlying Bell inequalities, some deny that the inequalities indicate anything significant about either locality or realism [13, § 9.3.2][16,32,45,47,158–160].

*3.3. Bell Inequalities: Measurement*

Dealing with measurement in quantum mechanics is not straightforward and requires significant revision to pre-quantum mechanical concepts.

> The result of a measurement is not ascribed to the systems, nor to their preparation, nor to the measurement, but to the totality. The totality is a closed phenomenon, and the prescribed probability distributions refer to this totality [161][162, § 6]. Measured values do not necessarily exist beforehand [13, § 4.6.1] ... If a property has not been measured, the formalism does not attribute any value [163, § 3]. Only one context justifies a claim that any member ... of the ensemble was originally such that a well-defined value could be attributed to the property being explored in the measurement. That context is when all single runs give the same outcome [164]. [3, § 4.1][6, § 26.4][21, p. 223]

Violation of Bell inequalities will also reflect, at least in part, these non straightforward aspects of measurement [140], aspects that would be evident even when dealing with a single non-composite system [158][165, § VI]. It has been suggested that the essential implication of violation relates to the measurement process, rather than the composite system itself [166]. At very least, the apparatus parameters, for different apparatus settings, need to be correctly taken into account [167]. Doing so (a) involves acknowledging the known contextuality of quantum mechanics (the fact that its prescribed probabilities are specific to a particular experimental context), and (b) may prevent Bell inequalities from being derived [13, § 9.1.3][16,134,168–173]

Other analysis also suggests that violation of Bell inequalities fundamentally relates to contextuality [174,175]. This analysis highlights the apparent assumption that there is a single probability space describing the all the statistical data relevant to Bell inequalities [32][33, § 5][36][85, § 8.5][174, § 6][175–179]. Regardless of whether the single probability space is assumed or derived [159,180], it is unlikely to be appropriate [3, § 1.8][35, p. 1603][82, pp. 10–11][181, § 1.6][182, p. 97][183, § 7], and so at least part of the reason for the violation of Bell inequalities might be that data from different probability spaces have been inappropriately combined [165, § 2][184–187]. In response to this concern:

- an alternative approach, involving joint simultaneous measurement, has recently been proposed [167,188]; and

- an alternative inequality, not requiring a single probability space, has been derived [87, pp. 83–85][189].

The nature and forms of contextuality are also subject to ongoing analysis [190–195].

*3.4. Bell Inequalities: Free Will*

> Some focus on the assumption that changing an apparatus setting does not affect the distribution of any variables that determine the measurement event outcomes (measurement independence or free will). Supporters of this free will assumption argue that correlations between the systems and the settings chosen would have to be amazingly strong for it to be violated [115,196]. This so-called conspiracy is, however, difficult to rule out [64, § 5.7.3][160, § 5][197][198][199, Appendix][200, § 4]. It is also consistent with the view that free will is only practical and epistemic [3, § 1.9]. Arguments for the free will assumption may themselves involve circular reasoning [201, § 5.1]. [3, § 5.5]

The idea that there is any clear link, between quantum mechanics and questions of determinism or free will, has been strongly challenged [3, § 1.9][202,203]. Despite the obvious need to carefully and adequately define what is meant by free will [204], frequent failure to do so contributes to confused thinking in this area [205].

Despite ongoing challenges to (what is referred to as) superdeterminism [206–208], the reasonableness of rejecting the free will assumption continues to be acknowledged [209, § V][210], explored [211,212], and defended [213–218][219, § 5].

*3.5. Bell Inequalities: Other Issues*

For some, Einstein-Podolsky-Rosen pairs and Bell inequalities suggest time symmetry [3, § 6.4][220–222], perhaps in combination with reverse causality [223, § 5.2][224–228], an adynamical spacetime [229], or a three dimensional timeless space [230].

Others suggest that Bell inequalities involve assumptions about distinguishability [231], ergodicity [232,233], time-independent variables [234] or temporal locality [223, § 7]. Challenging these assumptions can generate differing assessments of the implications of violation of the inequalities. Whether or not any assumptions necessarily imply counterfactual reasoning is not a straightforward question [235,236] and, either way, alternative approaches avoiding conterfactuals may be possible [237]. The potential need for many-valued logic [238,239] is also not straightforward.

Some suggest that violation of Bell inequalities reflects the need for an alternative approach to the concept of a composite system with two subsystems:

either more holistic [24, p. 7][240, § 7][241, § 6][242], or giving more emphasis to the statistical ensemble than its members [21, § 9.3.5]. This highlights the pervasive need in quantum mechanics to accept that pre-quantum mechanical concepts may no longer apply [24, p. 5][243, § 4.2].

There are significant challenges in appropriately analysing actual sequential-in-time measurements in quantum mechanics [35][244, § V][245]. These challenges are particularly relevant to relating theoretical analysis of Bell inequalities to practical experiments [82, pp. 10–11][246, § 1, § 6][247, § 5][248, § 3], and partly reflect more general challenges of fully analysing a realistic measurement apparatus and realistically modelling laboratory experiments [28,164]. In experiments aiming to demonstrate violation of Bell inequalities, problems can also arise relating to statistical analyses and data, either of which may be incomplete or incorrect [3, § 5.6][249–251].

It is not possible in practice to fully overcome the multiple challenges. Thus, there is ongoing and significant dissent [252] from the widespread view that a series of experiments in the last decade [18,253–255] has dealt with all the significant challenges simultaneously and adequately. Whether or not future Bell experiments succeed to a greater extent in overcoming the challenges, there will remain multiple potential implications of any results, depending what view is taken on the theoretical aspects of the relevant inequalities [16,131,233,252,256–258].

## 4. Conclusions (Features Emerging from the Map)

In order to appropriately assess the implications of violation of Bell inequalities, it helps to set them in the wider context of quantum mechanics.

- Appropriate analysis of probability is central to quantum mechanics, and crucial to assessing the implications of violation of Bell inequalities. Bell inequalities reflect the Boole inequality for joint probability distributions. (Sect. 2.1 above)

- Generally in quantum mechanics, statistical balance characterizes a series of measurement event outcomes, none of which are explicable by reference to pre-measurement properties. It is not yet clear what this implies about any physical reality underlying two-measurement-type statistical-balance-in-time within an ensemble of simple systems (Sect. 2.2 above).

- It seems hard, therefore, to find any definite implications about any physical reality underlying two-measurement-type

statistical-balance-in-space-and-time in an ensemble of composite systems (Sect. 2.3 above).

- The Einstein-Podolsky-Rosen thought experiment highlights that, in quantum mechanics, some pairs of systems must be treated as a single composite system. Bell suggested that some possible non-quantum mechanical treatments of such pairs would satisfy an inequality, but showed that the same inequality is violated by the predictions of quantum mechanics (Sect. 2.4 above).

- Confusion can arise in discussion of Bell inequalities from failing to apply the caution which should be motivated by the statistical balance which pervades quantum mechanics (Sect. 3.1 above).

There are several possible ways to assess the implications of violation of Bell inequalities. These can lead to differing conclusions.

- Violation of Bell inequalities constrains prequantum theories, but does not rule out the possibility that such a theory might reconcile some form of locality and realism to quantum mechanical predictions (Sect. 3.2 above).

- Violation of Bell inequalities might, at least partly, reflect (a) the known contextuality of quantum mechanics (the fact that its prescribed probabilities are specific to a particular experimental context) and (b) that data from different probability spaces have been inappropriately combined (Sect. 3.3 above).

- Bell inequalities assume that changing an apparatus setting does not affect the distribution of any variables that determine the measurement event outcomes (measurement independence or free will). The reasonableness of rejecting this free will assumption continues to be defended. (Sect. 3.4 above)

- Other suggestions are based on alternative approaches to time, distinguishability, ergodicity, logic, counterfactuals and ensembles. Again, many of these would reconcile at least one understanding of locality and realism to quantum mechanical predictions. It is also possible that experiments in the last decade may not have dealt with the significant challenges adequately. (Sect. 3.5 above)

In summary, violation of Bell inequalities appears to have some implications for the nature of physical reality, but none of these are definite. Many claims that there are definite implications reflect one or more of:

- lack of precision in non-mathematical language (Sects. 1, 3.2 and 3.4 above);

- assumptions inappropriate in quantum mechanics (Sects. 3.2 and 3.3 above);

- inadequate treatment of measurement statistics (Sects. 3.1 and 3.3 above); and

- underlying philosophical assumptions (Sects. 1, 3.2 and 3.5 above).

## 5. Amendments to the 2019 Review Glossary

The 2019 review included a glossary of intended meanings for many elements of the non-mathematical language used [3, § 8].The intended meanings for such language in this article are the same as in the 2019 review, subject to the three additions and one amendment set out below.

The 2019 review did not include Bell inequalities in its glossary. The intended meaning for this article is:

> **Bell inequalities**    any of the inequalities classified in the 2014 review *Bell nonlocality* [74, Appendix].

The 2019 review glossary had no entry for *probability space*, despite the frequent use of this term in that review. The intended meaning of *probability space* for this article is:

> **probability space**    a set of mutually exclusive possible *events*, in a specified (actual or notional) experimental context, to which probabilities can be assigned such that it is certain that one (and only one) of the *events* in the set will occur [259, § 3][260, § 2.1]

(The qualification "actual or notional" allows for the fact that quantum mechanics applies to unobserved events [3, § 3.1][261, p. xiii]. The meaning of *event* is as given in the 2019 glossary.) It follows that the intended meaning of *probability distribution* for this article is:

> **probability distribution**    the probabilities assigned to possible *events*, in a specific *probability space*

The 2019 review meaning for *statistical ensemble* was potentially ambiguous: "a set of systems which can be treated as identical, such as those prepared in an identical way". The failure to further define the phrase "treated as identical" may have been unhelpful. The revised meaning of *statistical ensemble* for this article is:

> **statistical ensemble**    an actual or virtual set of *systems* which have been (actually or notionally) *prepared* in an identical way [164] (The qualification "actual or virtual" reflects the fact that, in some cases, the ensemble might have only one physical member and many mental copies [3, § 3.2][262, p. 308]. The qualification "actually or notionally" reflects the fact that quantum mechanics applies to spontaneous events [3, § 3.1][263–265], passively recorded, and to unobserved events [3, § 3.1][261, p. xiii]. The meanings of *system* and *prepared* are as in the 2019 glossary.)

### Acknowledgements

I acknowledge the work of all the authors of the cited references. Their work has made mine possible. I also thank those authors with whom I have exchanged emails for their encouragement, insight and constructive comments. These have, I hope, led to greater clarity in the present article.